\documentclass[final,5p,times,twocolumn]{article}

\usepackage[utf8]{inputenc}
\usepackage{natbib}
\usepackage{graphicx}
\usepackage[margin=0.5in]{geometry}

\usepackage{lineno,hyperref}
\modulolinenumbers[5]

\usepackage{epstopdf}
\usepackage{verbatim}
\usepackage{mathrsfs}
\usepackage{array}

\newcolumntype{H}{>{\setbox0=\hbox\bgroup}c<{\egroup}@{}}

\begin{document}

\title{Reproducible empirical evidence of cosmological-scale asymmetry in galaxy spin directions: comment on arXiv: 2404.06617}


\date{}

\author{Lior Shamir\footnote{lshamir@mtu.edu}  \\ Kansas State University \\ 1701 Platt St \\ Manhattan, KS 66506, USA}

\maketitle

\begin{abstract}

The distribution of the spin directions of galaxies has been a question in the past decade, with numerous Earth-based and space-based experiments showing that the distribution is not necessarily random. These experiments were based on different statistical methods, one of them was a simple and empirically verified open source $\chi^2$ method. Patel \& Desmond (2024) proposed that previous experiments showing non-random distribution are flawed since they assume Gaussian distribution. To address that, they apply a new complex ad-hoc statistical method to several datasets, none of them except for one were used in the past to claim for a dipole axis. The new method showed that all datasets except for one exhibit isotropy. This paper discusses the soundness of the contention that Gaussian distribution cannot be assumed for galaxy spin directions. More importantly, simple empirical analyses show that the new statistical method is not fully responsive to asymmetry in the distribution of galaxy spin directions, and does not identify non-random distribution even in situations where a dipole axis clearly exists in the data, or when an artificial bias is added to the data to create an extremely non-random dataset. Results using Monte Carlo simulation show substantial differences between the results of the simulation and the results of the new statistical method. Code and data to reproduce the experiments are available, and released in a manner that is easily reproducible. Possible reasons leading to the results are also discussed. The claims that the actual results are different from the results reported in the papers are examined in an open and transparent manner. These claims are found to be inaccurate, as the previous literature results are fully reproducible. These findings further reinforce the need to study astrophysical or cosmological explanations for the non-random distribution.

\end{abstract}


\section{Introduction}
\label{introduction}

The nature of the distribution of galaxy spin directions as seen from Earth has been a matter of research in the past five decades, providing conflicting observations. Some observations using several different space-based \citep{shamir2020pasa,shamir2021aas,Shamir_2024} and Earth-based \citep{macgillivray1985anisotropy,longo2011detection,shamir2012handedness,shamir2016asymmetry,shamir2019large,shamir2020patterns,shamir2020large,shamir2020pasa,shamir2021particles,shamir2021large,shamir2022new,shamir2022large,shamir2022asymmetry,shamir2022analysis,Shamir_2024} instruments used large numbers of galaxies and consistently showed that the distribution of galaxy spin directions in the Universe is not random, and exhibits a dipole axis formed by the galaxy spin directions as observed from Earth. Most recently, JWST showed that the asymmetry in the distribution of galaxy spin direction can be observed with a small number of galaxies, and therefore without the need for sophisticated statistical analysis \citep{Shamir_2024}.

On the other hand, several other studies argued that the directions of rotation of spiral galaxies are random \citep{iye1991catalog,land2008galaxy,hayes2017nature,tadaki2020spin,iye2020spin,jia2023galaxy}. Detailed analysis of all of these experiments showed that in all cases the data in fact showed non-random distribution \citep{sym15091704}. In some cases the reasons were experimental design decisions mentioned in the papers, and in other cases a full reproduction was needed since these decisions were not mentioned in the paper, and were only revealed after contacting the research institutions where the research was conducted. Details of the analysis, as well as code and data in the cases full reproductions were needed are described in \cite{sym15091704}, as well as in dedicated sections in \citep{psac058,mcadam2023reanalysis,shamir2022analysis,Shamir_2024}.

The analysis of the distribution of galaxy spin directions was done using several different methods. Most of the previous work used simple binomial distribution, as well as $\chi^2$ statistics, such as in \citep{shamir2020pasa,shamir2021particles,shamir2020patterns,shamir2022analysis,shamir2021large,shamir2022new,shamir2022large}. These methods were tested under different conditions using control experiments and synthetic data to fully profile the way they respond to different possible anomalies in the distribution of galaxy spin directions \citep{shamir2021particles}. Using synthetic data allowed to create certain controlled conditions, and test the ability of the statistics to identify them \citep{shamir2021particles}, ensuring that the methodology is sound. Monte Carlo simulations were also used \citep{psac058} to verify that the statistics is consistent. 

When the size of the galaxy databases became larger, an analysis was done without any attempt to fit the galaxies into a pre-defined model. Instead, a direct comparison of the number of clockwise and counterclockwise galaxies around each point in the sky was done in \citep{shamir2022analysis}. Using the power of JWST, a very simple binomial distribution analysis showed a strong non-random distribution of galaxy spin directions in the JWST deep field taken at the HST Ultra Deep Field footprint. That analysis can even be done by manually looking at the image, and without the need for sophisticated algorithms or statistical models \citep{Shamir_2024}.

Recently, \cite{petal} argued that the distribution of galaxy spin directions is random, and the observed asymmetry and dipole axes were the results of an incorrect assumption that the probability of a spiral galaxy to rotate in a certain direction does not follow normal distribution. This paper examines the claims, investigates the analysis, and reproduces the results to evaluate the soundness, robustness, and reproducibility of the new statistical analysis proposed by \citep{petal}.

\section{Assuming normal distribution}
\label{normal_distribution}

Previous experiments were based on several different types of analyses, one of them is an analysis of the probability of a dipole axis to be formed by the distribution of the galaxy spin directions. That was done in \citep{longo2011detection,shamir2012handedness,shamir2019large,shamir2020patterns,shamir2020large,shamir2020pasa,shamir2021particles,shamir2021large,shamir2022new,shamir2022large,shamir2022asymmetry,shamir2022analysis} using simple $\chi^2$ statistics. 

In summary, the $\chi^2$ analysis was done by fitting the spin directions of the galaxies to the cosine of the angular distance between all galaxies in the dataset and every $(\alpha,\delta)$ combination in the sky. Equation~\ref{chi2} shows the $\chi^2$
\begin{equation}
\chi^2_{(\alpha,\delta)}=\Sigma_i | \frac{(d_i \cdot | \cos(\phi_i)| - \cos(\phi_i))^2}{\cos(\phi_i)} | ,
\label{chi2}
\end{equation}
where $d_i$ is 1 if galaxy {\it i} rotates clockwise or -1 if galaxy {\it i} rotates counterclockwise, and $\phi_i$ is the distance in degrees between galaxy {\it i} and the tested dipole axis $(\alpha,\delta)$.

The statistical significance of a dipole axis to exist at $(\alpha,\delta)$ is determined by assigning all galaxies with random spin directions, and computing the $\chi^2_{(\alpha,\delta)}$ a large number of times for each $(\alpha,\delta)$ combination. The $\sigma$ difference between the mean $\chi^2$ when $\chi^2_{(\alpha,\delta)}$ is computed with random spin directions and the $\chi^2_{(\alpha,\delta)}$ computed with the real spin directions determines the probability of such axis to form by chance. That is repeated for each $(\alpha,\delta)$ combination to shows the mere chance probability of a dipole axis to form at all parts of the sky.

The random spin direction $d_i$ of galaxy $i$ is determined simply by Equation~\ref{random}
\begin{equation}
d_i = rand<\{-1,1\}>.
\label{random}
\end{equation}

The motivation to use a random spin direction is that the spin directions, the position angles, the inclinations, and the positions of spiral galaxies are all expected to be distributed randomly. In that case, the probability of a galaxy to rotate in a certain direction as viewed from Earth would be exactly 0.5, regardless of its location in the Universe or any other condition. \cite{petal}, however, argue that the use of $\chi^2$ is flawed for the reason that the spin directions of galaxies do not follow Gaussian distribution. The argument is provided without a detailed explanation. 



To avoid assuming normal distribution, \cite{petal} proposed a new statistical method designed specifically for the purpose of identifying anisotropy in data of galaxy spin directions. The method is complex, goes through several steps and components, and therefore also difficult to reproduce from the information in the paper alone in a manner that secures that the implementation is identical to the code used by \cite{petal}. The method does not have available source code, and attempts to obtain the code through private communication were declined. On the positive side, the last author of \citep{petal} indicated they will consider releasing the code in the future. The last author of \citep{petal} also indicated that the project was done by an undergraduate summer student. While the level of experience of the experimentalist does not necessarily correlate with their ability to generate good code, the results shown in Section~\ref{reproducability} exhibit clear evidence of coding errors.

Unlike the analysis method used in \citep{psac058,mcadam2023reanalysis,shamir2022analysis,shamir2021large,shamir2022new,shamir2022large}, which was inspected by a large number of scientists and tested under multiple conditions using synthetic data to profile its behavior \citep{shamir2021particles}, the method of \cite{petal} is new, and has not yet been tested thoroughly. The method is complex, and has multiple steps , and its code is not released to the public. The inalienability of the code makes the method more difficult to examine and verify its correctness. 

A very important note is that \cite{petal} necessarily assume that any dipole axis formed by the distribution of galaxy spin directions is a feature of the large-scale structure of the Universe. However, previous observations have provided evidence that the photometry of a galaxy depends on its direction of rotation \citep{hamir2013color,shamir2016asymmetry,sym14050934}. More specifically, the observations showed that the photometry and spectroscopy of galaxies that rotate in the same direction relative to the Milky Way are different from the photometry and spectroscopy of galaxies that rotate in the opposite direction relative to the Milky Way \citep{shamir2017large,shamir2020asymmetry,sym15061190,shamir2024simple}. In that case, the rotational velocity might affect the light of the observed galaxies in a manner that is more substantial than expected \citep{sym15061190,shamir2024simple}.

\begin{figure*}
\centering
\includegraphics[scale=0.5]{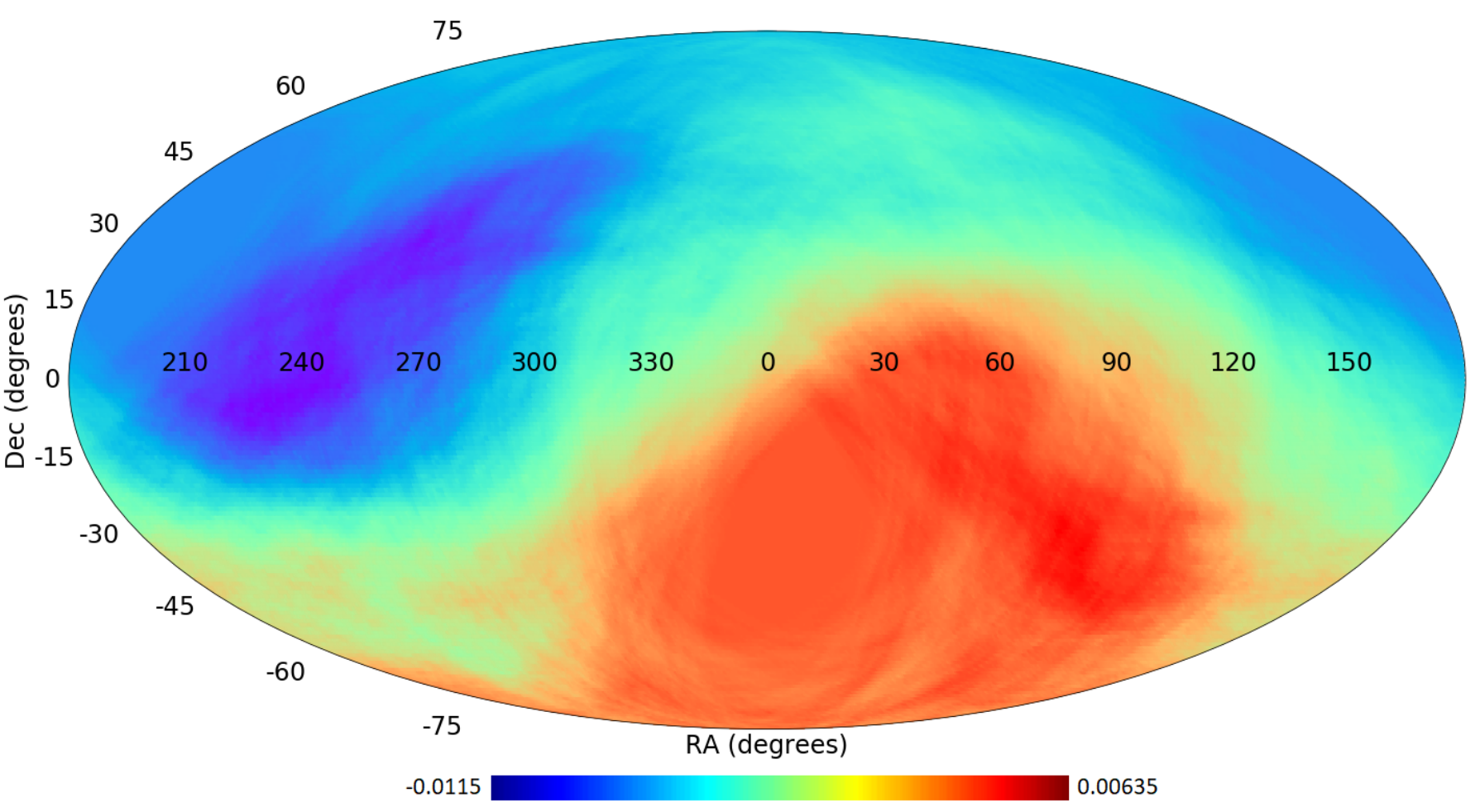}
\caption{Asymmetry between the number of galaxies spinning in different directions at different parts of the sky. The analysis was done with $\sim1.3\cdot10^6$ galaxies taken from both the Northern and Southern hemispheres of the DESI Legacy Survey. The measurement of the asymmetry is direct to each point in the sky, and not the result of fitting the distribution into a certain model (e.g, dipole axis). The results show a dipole axis formed at close proximity to both ends of the galactic pole. Because the analysis is a direct measurement and not fitting to a statistical model, the asymmetry in one hemisphere is not expected to affect the asymmetry in the opposite hemisphere \citep{shamir2022analysis}.}
\label{all10_6K_dist90_inc1}
\end{figure*}

If galaxies located around the galactic and pole rotate in a certain direction are brighter than the galaxies in the same footprint that rotate in the opposite direction \citep{shamir2017large,shamir2020asymmetry,sym15061190}, more of the brighter galaxies will be observed from Earth. Since these brighter galaxies have a spin pattern, that will lead an Earth-based observer to believe that there is an asymmetry between the number of galaxies with opposite spin directions. The opposite end of the galactic pole will show inverse asymmetry, leading to a dipole axis that is not driven by the large-scale structure of the Universe, but by the impact of the rotational velocity on the brightness of the galaxies. Indeed, most dipole axes reported so far peak with close proximity to the galactic pole, as also shown by Figure 7 in \citep{sym15091704}. Also, analysis of $\sim1.3\cdot10^6$ galaxies that cover both the Southern and Northern hemispheres that was done without fitting to a statistical model showed a distribution of asymmetry aligned with the two ends of the galactic pole \citep{shamir2022analysis}. This can be seen in Figure~\ref{all10_6K_dist90_inc1}, that was generated by direct measurements of the asymmetry around each point in the sky, and without fitting into any statistical model of the Universe \citep{shamir2022analysis}. 

\section{Empirical testing}
\label{empirical}

A dipole axis formed by the distribution of galaxy spin directions means that the sky can be separated into two hemispheres such that one hemisphere has an excessive number of galaxies rotating clockwise, while the opposite hemisphere has a higher number of galaxies rotating counterclockwise. This analysis is obviously simple, but its advantage in the sense of this study is that it is a  straightforward binomial statistics that is easy to understand, and the experimentalist can easily see the distributions and their statistical significance through the data. That analysis does not require complex methods that run through ad-hoc code, and can be therefore considered more robust and less vulnerable to possible coding or statistics errors. These experiments were done in all papers used in \cite{petal}, although these experiments are not mentioned in \citep{petal}. 

Another advantage of separating the galaxies into two hemispheres with inverse spin direction distribution asymmetry is that it allows very easy analysis through Monte Carlo simulations, allowing to determine the probability of the alignment to occur by chance without making assumptions driven by the selection of a certain statistical method. The simplicity and robustness of the analysis can therefore be used to test ad-hoc statistical methods.

One of the datasets used by \cite{petal} is the dataset of 72,888 galaxies used in \citep{psac058,sym15091704}. \cite{petal} mistakenly credit this author for that dataset, but that dataset was compiled by \cite{iye2020spin} from a catalog of photometric objects \citep{shamir2017photometric}, and provided to this author by Dr. Brad Gibson through private communication. As stated explicitly in Section 9 of \citep{psac058}, ``the dataset used here does not necessarily prove that the distribution of the spin directions of SDSS galaxies forms a dipole axis''. It was analyzed to test whether it can disprove such claim, and was found consistent with results made by the larger datasets that are more appropriate for this task.

As also stated in \citep{psac058}, the dataset was not prepared for the purpose of identifying a dipole axis in the distribution of galaxy spin directions, and the galaxies were annotated based on just 10 peaks in its radial intensity plot as described in \citep{shamir2017photometric}. That leads to some inaccuracies in the annotation of the galaxy spin directions while increasing the number of galaxies that meet the threshold for annotation. That is different from the catalogs annotated with a minimum of 30 peaks in the radial intensity plot, such as in the large catalogs used to study the distribution of galaxy spin directions in the Universe \citep{shamir2020patterns,shamir2020large,shamir2020pasa,shamir2021particles,shamir2021large,shamir2022new,shamir2022large,shamir2022asymmetry,shamir2022analysis}. Examples of such data are available at several sources such as \citep{shamir2024simple}. As shown theoretically and empirically, inaccuracies in the annotation of the galaxies can weaken the statistical signal but cannot increase it, unless the error is not symmetric \citep{shamir2021particles}.

Table~\ref{hemispheres1} shows the number of galaxies that rotate in opposite directions in two opposite hemispheres. As the table shows, according top the data the sky can be separated into two hemispheres, such that in the hemisphere $(70 < \alpha < 250)$ there is a higher number of galaxies rotating clockwise, and in the opposite hemisphere there is a higher number of galaxies rotating counterclockwise. The data is taken from the open dataset that can be accessed at \url{https://people.cs.ksu.edu/~lshamir/data/iye_et_al}, and discussed in \citep{psac058,sym15091704}, Obviously, the analysis is extremely simple, and can be reproduced easily.

By taking the conservative two-tailed binomial probability, and when assuming that in the hemisphere $(\alpha >250 \cup \alpha <70 )$ the distribution is random, the probability to have such asymmetry or stronger to occur by chance is $p=\sim0.014$. Although the analysis is simple, it was also verified by a Monte Carlo simulation as described in \citep{psac058}, providing a probability that the sky can be separated into two hemispheres with such asymmetry or stronger of $p=\sim0.008$. The p value of the Monte Carlo simulation is somewhat smaller than the P value of the binomial distribution, which can be naturally explained by the fact that the binomial distribution analysis conservatively assumes random distribution in the hemisphere $>250 \cup <70$. Code and instructions to reproduce the results are available \footnote{\url{https://people.cs.ksu.edu/~lshamir/data/iye_et_al}}.

The Monte Carlo simulation is run 100,000 times, such that in each run the galaxies are assigned with random spin directions, and the number of times an asymmetry equal or stronger to the asymmetry shown in Table~\ref{hemispheres1} is counted. It is important that the Monte Carlo simulation does not test the probability of such asymmetry or stronger to occur by chance in the one selected hemisphere separation shown in Table~\ref{hemispheres1}, but to occur by chance in any two hemispheres in the sky. That is, in each run of the simulation the distribution was tested whether it satisfies such distribution in any two hemispheres. That ensures that the use of different hemispheres is accounted for, although the separation into two hemispheres does not provide independent variables, and therefore the binomial distribution analysis is valid. In any case, the Monte Carlo simulation ensures that the distribution is not random.

\begin{table}
\scriptsize
\begin{tabular}{lccccc}
\hline
Hemisphere & \# cw  & \# ccw  & $\frac{cw}{ccw}$   & P               & P   \\
(RA)           &            &               &                          & (one-tailed)  &  (two-tailed)  \\
\hline
70$^o$-250$^o$ & 23,037 & 22,442 & 1.0265 & 0.0026 & 0.0052  \\
$>250^o \cup <70^o$ & 13,660 & 13,749 & 0.9935 & 0.29 & 0.58  \\
\hline
\end{tabular}
\caption{The number of galaxies rotating in opposite directions using dataset ``Shamir (a)'' in \citep{petal}, and the binomial probability of the asymmetry to occur by chance in each hemisphere. The analysis is discussed thoroughly in \citep{psac058,sym15091704}, and the data are publicly available at \url{https://people.cs.ksu.edu/~lshamir/data/iye_et_al}. This table is also Table 2 in \citep{sym15091704}.}
\label{hemispheres1}
\end{table}

\begin{figure}
\centering
\includegraphics[scale=0.27]{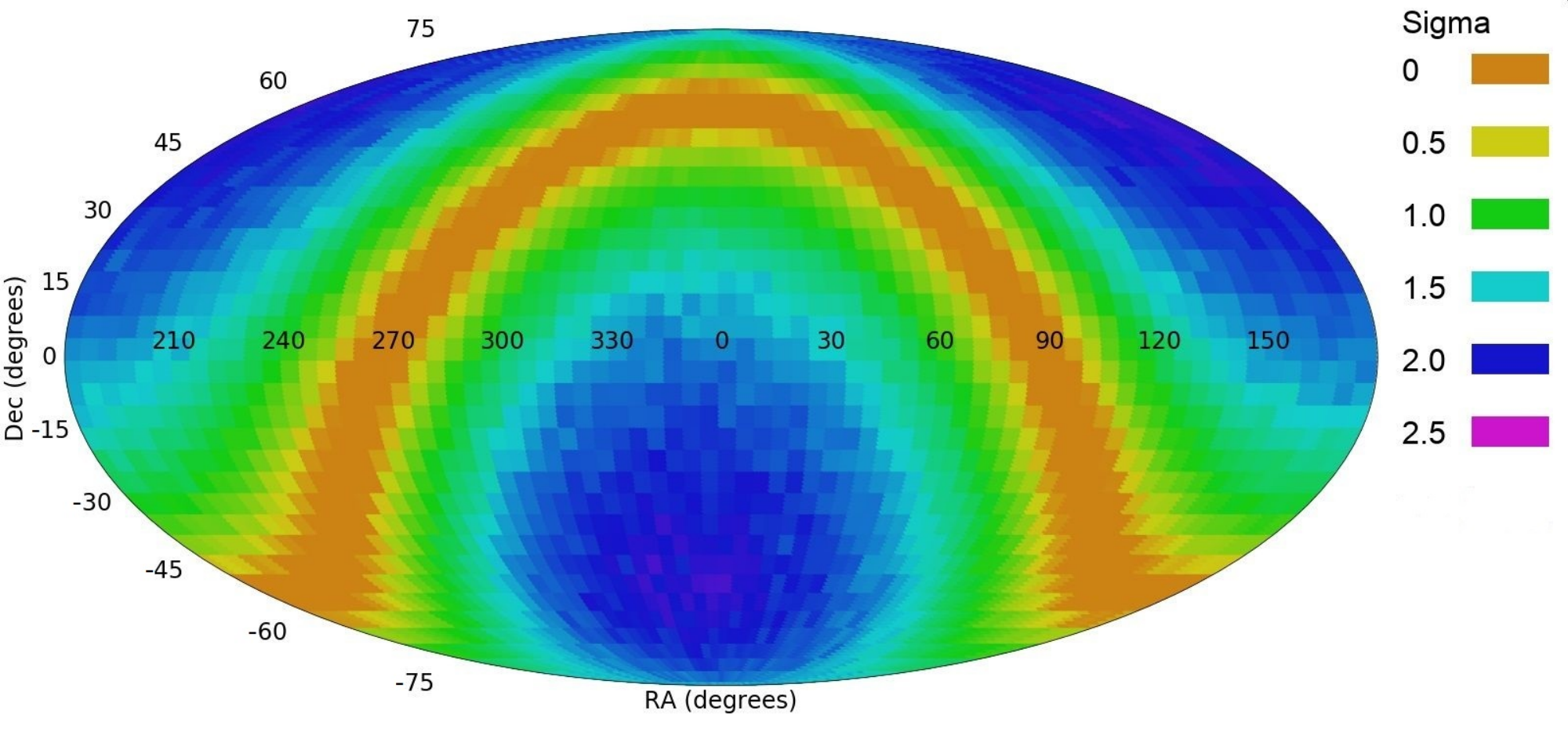}
\caption{The statistical significance of a dipole axis from all $(\alpha,\delta)$ combinations. Full reproduction is available at \url{https://people.cs.ksu.edu/~lshamir/data/iye_et_al}. That is dataset ``Shamir (a)'' in \citep{petal}}
\label{dipole1}
\end{figure}

The observation that the sky can be separated into two hemispheres with opposite asymmetry in the distribution of the directions of rotation indicates that the dataset fits a statistically significant dipole axis. Indeed, the analysis in \citep{psac058,sym15091704} shows a dipole axis with probability of 2.12$\sigma$ that peaks at $(\alpha=170^o,\delta=35^o)$. The code and data to reproduce the analysis is available at \url{https://people.cs.ksu.edu/~lshamir/data/iye_et_al}. Figure~\ref{dipole1} visualizes the results of the analysis.

As Table 4 in \citep{petal} shows, the statistical method used by \cite{petal} provides a probability of 0.32 to have a dipole axis by chance. That is in disagreement with the data shown in Table~\ref{hemispheres1}. The table shows in a simple manner that the sky can be separated into two hemispheres such that one has a higher number of clockwise galaxies and the other has a higher number of counterclockwise galaxies. Straightforward binomial distribution analysis as well a Monte Carlo simulation both show that it is unlikely to happen by chance, yet the statistical method used by \cite{petal} did not detect that anomaly. Therefore, while the data shows two hemispheres with inverse asymmetry in their spin directions that is unlikely to happen by chance, the statistical method used by \cite{petal} does not identify it. That is, the data show that the sky can be separated into two hemispheres that are different from each other, and binomial distribution and Monte Carlo simulation both show that this separation is unlikely to happen by chance, and yet the method of \cite{petal} shows that the Universe as reflected by these galaxies is isotropic. 

The method of \cite{petal} does not have code, and therefore it cannot be inspected, and it is difficult to know what caused the method to miss a clear difference between the two hemispheres that is expect to exhibit itself in the form of a dipole axis. The complex nature of the ad-hoc statistical method makes it more difficult to analyze the method, and therefore the open source allowing to know how the method works might be an important tool to analyze the behavior of the method. Also, unlike the analysis in \citep{psac058,sym15091704}, the method of \cite{petal} does not provide the location of the axes and the statistical strengths of the possible axes in the sky, which helps to understand how the method performs. The limited output combined with the absence of the source code makes it difficult to assess the statistical method used by \cite{petal}. 

Similarly, dataset ``Shamir (b)'' in \citep{petal} can also be separated into two hemispheres, with stronger statistical signal compared to ``Shamir (a)''. The separation can be reproduced easily using the data, available at \url{http://people.cs.ksu.edu/~lshamir/data/assymdup}. Table~\ref{hemispheres2} shows the distribution of the spin directions of the galaxies in the two hemispheres.

\begin{table}
\scriptsize
\begin{tabular}{lccccc}
\hline
Hemisphere & \# cw  & \# ccw  & $\frac{cw}{ccw}$   & P               & P   \\
(RA)           &            &               &                          & (one-tailed)  &  (two-tailed)  \\
\hline
                $70^o-250^o$         & 24,648 & 23,958 & 1.0288 & 0.00086 & 0.00172  \\
              $>250^o \cup <70^o$ & 14,540 & 14,694 & 0.989 & 0.1823     & 0.364  \\
\hline
\end{tabular}
\caption{The number of galaxies rotating in opposite directions in two opposite hemispheres using dataset ``Shamir (b)'' in \citep{petal}. The data are publicly available at \url{http://people.cs.ksu.edu/~lshamir/data/assymdup}.}
\label{hemispheres2}
\end{table}

\begin{figure}[h]
\centering
\includegraphics[scale=0.25]{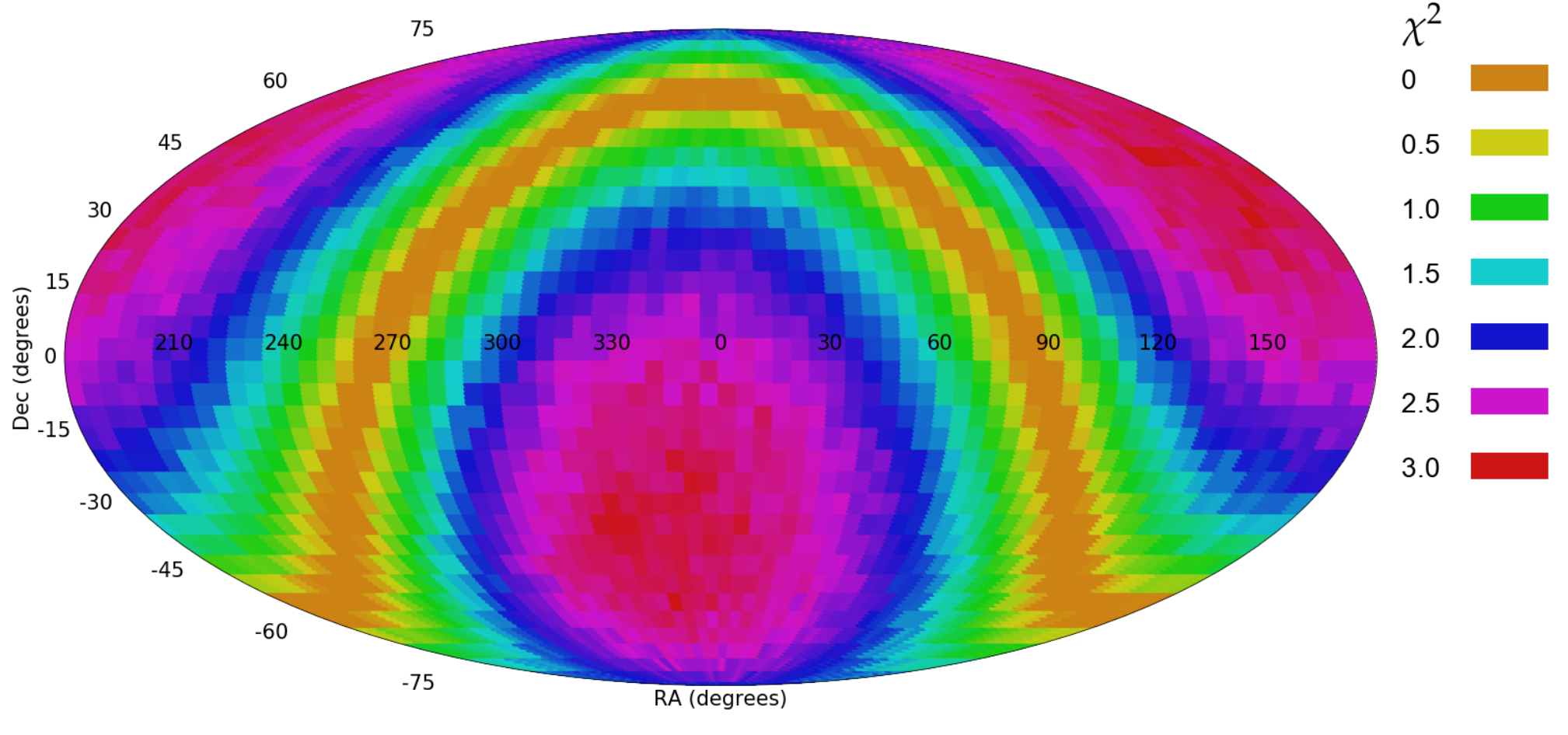}
\caption{Statistical significance of a dipole axis from all different $(\alpha,\delta)$ integer combinations. The data are taken from \citep{shamir2021particles}, and labeled as ``Shamir (b)'' in \citep{petal}.}
\label{dipole2}
\end{figure}

Even if assuming that the spin directions in the hemisphere $(>250^o \cup <70^o)$ are distributed randomly, the probability of such distribution to occur by chance is 0.0034. A Monte Carlo simulation to have such asymmetry or stronger in any two hemispheres by chance was 0.0019. Again, the fact that one hemisphere has a significantly higher number of galaxies spinning clockwise and the opposite hemisphere has a higher number of galaxies spinning counterclockwise indicates that the galaxy spin directions form a dipole axis. Therefore, a statistical method that can identify a dipole axis must be able to identify such alignment. The method used in \citep{shamir2021particles} as well as the other previous experiments provides the results visualized by Figure~\ref{dipole2}. As the figure shows, the method was able to detect a dipole axis that peaks at $(\alpha=165^o,\delta=40^o)$, with statistical significance of 2.56$\sigma$. 

The method used by \cite{petal} shows a P value of 0.32. That shows that the method is not able to identify the dipole axis formed by the two hemispheres with inverse asymmetry in galaxy spin direction. It is also surprising that the P values provided by the  \cite{petal} statistical method is the same for the dataset in Table~\ref{hemispheres1} and the dataset in Table~\ref{hemispheres2}. The statistical significance computed through the $\chi^2$ analyses for the dataset in Table~\ref{hemispheres2} is different from the one in the dataset of Table~\ref{hemispheres1}, ``Shamir (a)'' in the \cite{petal} notation. Also, the binomial distribution P value of ``Shamir (a)'' is 0.014 while the P value of ``Shamir (b)'' (Table~\ref{hemispheres2}) is 0.0034. So while these datasets are similar in size and footprint, and might therefore be expected to be similar, simple statistics shows that they are still different by the distribution of their data. It can therefore be considered surprising that the statistical method of \cite{petal} assigns them with the same probability.

Perhaps the most relevant dataset in the sense of studying the statistical method of \cite{petal} is the mirrored dataset of the galaxies annotated by {\it SpArcFiRe}. The {\it SpArcFiRe} algorithm is known to have a consistent bias in its annotation \citep{hayes2017nature}. Unlike the ``mechanical'' nature of {\it Ganalyzer}, the {\it SpArcFiRe} is far more sophisticated, and therefore its code is more difficult to follow and correct. \cite{hayes2017nature} reported on the bias and acknowledged they were not able to fully correct it, as described in the appendix of \citep{hayes2017nature}. That bias is also noticeable empirically by the dramatic differences between the results observed with the original images and the results observed when the galaxy images are mirrored \citep{mcadam2023reanalysis}. 

That means that even if the spin directions are distributed randomly in the Universe, a dataset annotated by {\it SpArcFiRe} should detect a non-random distribution added artificially by the algorithm. That is, when using galaxies annotated by {\it SpArcFiRe}, and given a sufficient number of galaxies, the null hypothesis is an abnormal Universe in which galaxy spin directions are not distribute randomly. {\it SpArcFiRe} also has a certain degree of inaccuracy. The reason it was used in the past was not necessarily to study the Universe, but to reproduce the results of \cite{hayes2017nature} after eliminating the step of manually removing features that correlate with the galaxy spin directions \citep{mcadam2023reanalysis}. That is, the dataset was not used to prove the existence of a dipole axis, but to show that the analysis of \cite{hayes2017nature} cannot disprove it.

That dataset is based on SDSS galaxies with spectra, and therefore its footprint is somewhat different from the other two datasets discussed above. But although it has not been a primary dataset used to study the distribution of spin direction, it still shows two hemispheres such that one of them has a higher number of galaxies spinning clockwise, and the opposite hemisphere has a higher number of galaxies spinning counterclockwise. Since the asymmetry is statistically significant according to simple binomial distribution it is expected that a statistical method would detect it. As mentioned above, {\it SpArcFiRe} is known and verified to be a biased algorithm, and therefore data annotated by it is expected in any case to lead to non-random distribution if the dataset is sufficiently large, regardless of the properties of the Universe. Previous control experiments with synthetic data showed that even a small artificial bias in the data leads to a dipole axis with an extremely strong statistical significant \citep{shamir2021particles}. 

\begin{figure}[h]
\centering
\includegraphics[scale=0.17]{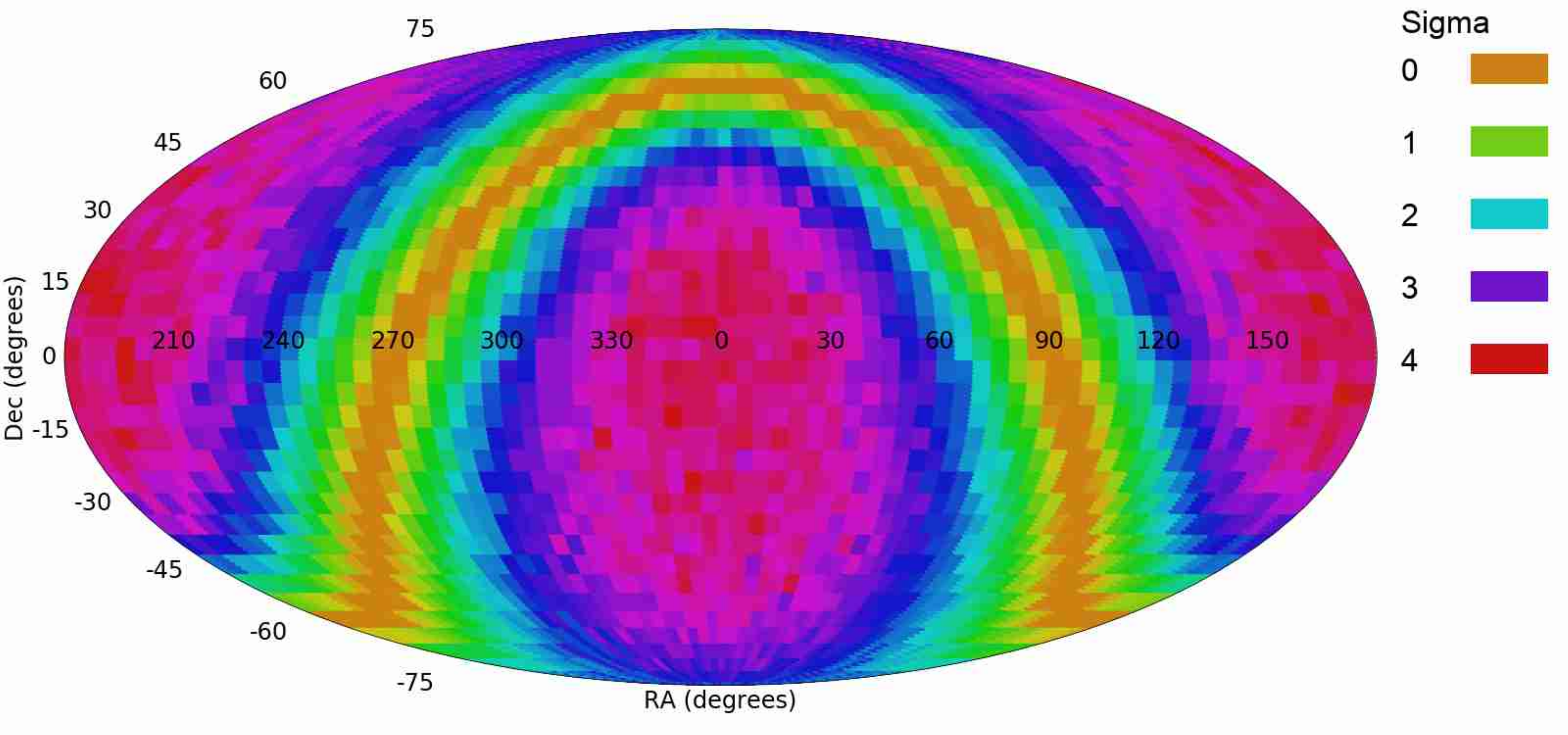}
\caption{Statistical significance of a dipole axis from all different $(\alpha,\delta)$ integer combinations using the dataset of mirrored galaxy image annotated by the {\it SpArcFiRe} algorithm. The data are taken from \citep{mcadam2023reanalysis}, and labeled as ``GM M'' in \citep{petal}.}
\label{sparcfire2}
\end{figure}

While the {\it SpArcFiRe} bias is difficult to profile, the only relevant information for the statistical analysis is the outcome of the annotations, and the distribution of the data. Table~\ref{sparcfire} shows the distribution of clockwise and counterclockwise galaxies in opposite hemispheres. Figure~\ref{sparcfire2} shows the probability of a dipole axis of galaxy spin directions to be formed in the data by chance. While the known {\it SpArcFiRe} bias might be small, the data itself as shown in Table~\ref{sparcfire} is distributed in an extremely non-random manner, with very strong P value for the asymmetry.

\begin{table}
\scriptsize
\begin{tabular}{lccccc}
\hline
Hemisphere & \# cw  & \# ccw  & $\frac{cw}{ccw}$   & P               & P   \\
(RA)           &            &               &                          & (one-tailed)  &  (two-tailed)  \\
\hline
                $120-300$         & 62,403 & 60,887 & 1.0249 & $<10^{-5}$ & $<10^{-5}$  \\
              $>300 \cup <120$ & 8,269 & 8,293 & 0.9971 & 0.423  &  0.826  \\
\hline
\end{tabular}
\caption{The number of galaxies rotating in opposite directions using dataset ``GAN M'' in \citep{petal}. The data are publicly available in \url{https://people.cs.ksu.edu/~lshamir/data/sparcfire/}.}
\label{sparcfire}
\end{table}

The distribution of the spin directions of the galaxies in the dataset is extremely non-random, as can be expected regardless of the properties of the Universe due to the {\it SpArcFiRe} internal bias. A statistical method that analyzes the data must provide results that the galaxy spin directions in the universe, as annotated by {\it SpArcFiRe}, are not randomly distributed. A statistical analysis that examines such highly asymmetric dataset and concludes that the distribution is random is limited in its ability to detect anomalies in galaxy spin directions. The p values of 0.25 as shown for that dataset by \cite{petal} suggests that the method is not sensitive to non-random distribution. As mentioned above, whether this specific dataset is of astronomical meaning is questionable due to the asymmetric nature of the {\it SpArcFiRe} method, but that is a topic unrelated to the statistical analysis. The dataset itself is not nearly random. As in the other datasets, opposite hemispheres show inverse asymmetries. But even when not separating to hemispheres, the simple binomial probability that the dataset is random is $p<6\cdot10^{-5}$. Separation to two hemispheres increases the signal, as can be expected since the asymmetry in one hemisphere is inverse to the asymmetry in the opposite hemisphere. A method that analyzes that dataset and concludes that the distribution of galaxy spin directions in it is random is limited in its ability to identify anomalies in the distribution of galaxy spin directions.

\section{Reproducibility of previous experiments}
\label{reproducability}

\cite{petal} also attempted to reproduce three previous experiments. The reproduction is summarized in Section 4.3 in \citep{petal}, titled ``comparison to literature''. The experiments they reproduce include the two datasets annotated by {\it SpArcFiRe} and used in \citep{mcadam2023reanalysis}, and the dataset used by \cite{longo2011detection}. For all three experiments they argue that the results of their reproduction do not agree with the results reported in the papers.  \cite{petal} do not provide information about the results of their reproduction, but broadly claim that their results are different from the results stated in the papers that describe the original experiments. No further information is provided regarding the experiments or the results. 

Despite the claims made by \cite{petal}, all three experiments are fully reproducible to the level reported in the papers, and there is no difficulty in producing the exact same results as reported. Code, data, and instructions to reproduce the results are available at \url{https://people.cs.ksu.edu/~lshamir/data/patel_desmond/}. The reproduction is fully transparent, but also provides very simple instructions to allow any interested person to reproduce all experiments in a convenient and accessible manner. 

In summary, the analysis of the dataset of mirrored galaxy images annotated by {\it SpArcFiRe} provided the results shown in Table~\ref{analysis1}. As the table shows, the results of the reproduction are virtually the same as the results reported in \citep{mcadam2023reanalysis}. The small differences are expected statistical fluctuations that are the natural results of result of simulations.  Code, data, and instructions to reproduce the results are available at \url{https://people.cs.ksu.edu/~lshamir/data/patel_desmond/}.

\begin{figure}[h]
\centering
\includegraphics[scale=0.27]{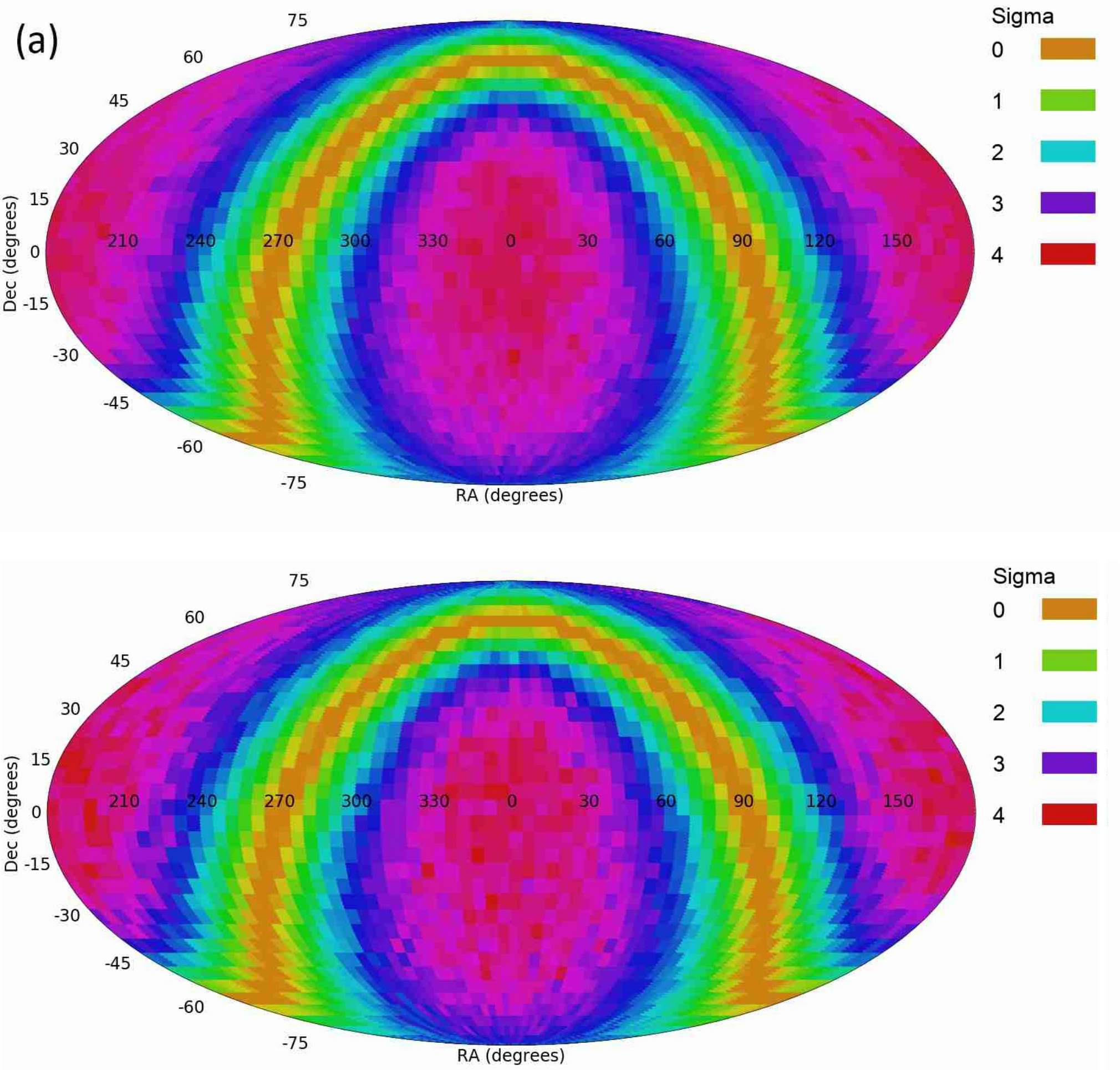}
\caption{Figure 2(a) in \citep{mcadam2023reanalysis} (top), and a visualization of the results of the reproduction using the {\it SpArcFiRe} annotated data used in \citep{mcadam2023reanalysis} available at \url{https://people.cs.ksu.edu/~lshamir/data/sparcfire} and the code available at \url{https://people.cs.ksu.edu/~lshamir/data/patel_desmond/}. The two figures are virtually identical, showing high level of reproducibility.}
\label{paper_reproduction}
\end{figure}

Figure~\ref{paper_reproduction} shows a comparison between the reproduction using the {\it SpArcFiRe} annotated data used in \citep{mcadam2023reanalysis} and available at \url{https://people.cs.ksu.edu/~lshamir/data/sparcfire}, and Figure 2(a) in  \citep{mcadam2023reanalysis}. The comparison shows that the two figures are virtually identical.

\begin{table}
\begin{tabular}{lcc}
\hline
    &  $(\sigma)$                       &      Location of maximum axis   \\
\hline
Reproduction	                       & 4.02	& $(\alpha=185^o, \delta=15^o)$ \\
\citep{mcadam2023reanalysis}  &	3.97	& $(\alpha=184^o, \delta=16^o)$ \\
\hline
\end{tabular}
\caption{The results of the reproduction and the values specified in \citep{mcadam2023reanalysis} computed with the dataset of mirrored galaxy images annotated by {\it SpArcFiRe}, which are also the values mentioned in \cite{petal}. The numbers can be reproduced with code and data from \url{https://people.cs.ksu.edu/~lshamir/data/patel_desmond/}.}
\label{analysis1}
\end{table}

\begin{figure}[h]
\centering
\includegraphics[scale=0.16]{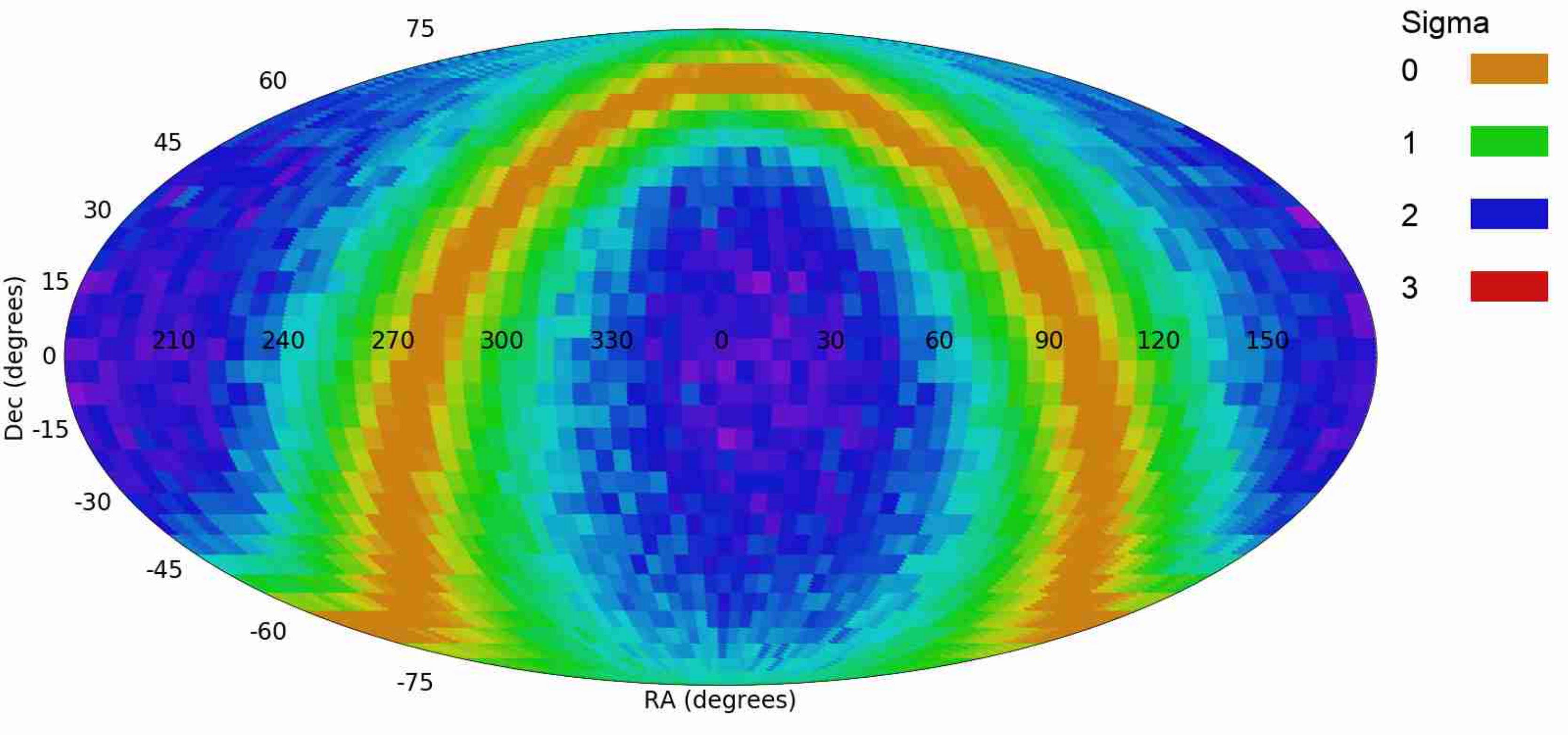}
\caption{Statistical signal in each part of the sky for a dipole axis formed by the distribution of galaxy spin directions in the non-mirrored galaxies annotated by {\it SpArcFiRe} \citep{mcadam2023reanalysis}.}
\label{ganalyzer_non_mirrored_4d_results}
\end{figure}

The same is with the dataset of non-mirrored galaxy images annotated by {\it SpArcFiRe}. Table~\ref{analysis2} shows the comparison between the results of the reproduction and the results reported in \citep{mcadam2023reanalysis}.  As before, the results of the reproduction are basically identical to the results reporte in \citep{mcadam2023reanalysis}. Figure~\ref{ganalyzer_non_mirrored_4d_results} shows the all-sky projection of the statistical significance of a dipole axis to form from each part of the sky. The dataset did not have a figure in \citep{mcadam2023reanalysis} so no comparison to the paper is possible, but it is clear that the numbers reported in \citep{mcadam2023reanalysis} can be easily reproduced from the data to a very high level of similarity. It is therefore unclear why \cite{petal} were not able to reproduce the results.

\begin{table}
\begin{tabular}{lcc}
\hline
    &  $(\sigma)$                       &      Location of maximum axis   \\
\hline
Reproduction	                       & 2.34 & $(\alpha=190^o, \delta=20^o)$ \\
\citep{mcadam2023reanalysis}  & 2.33 & $(\alpha=192^o, \delta=24^o)$ \\
\hline
\end{tabular}
\caption{The results of the reproduction and the values specified in \citep{mcadam2023reanalysis} computed with the dataset of non-mirrored galaxy images annotated by {\it SpArcFiRe}. The numbers can be reproduced with code and data from \url{https://people.cs.ksu.edu/~lshamir/data/patel_desmond/}.}
\label{analysis2}
\end{table}

\begin{figure}[h]
\centering
\includegraphics[scale=0.16]{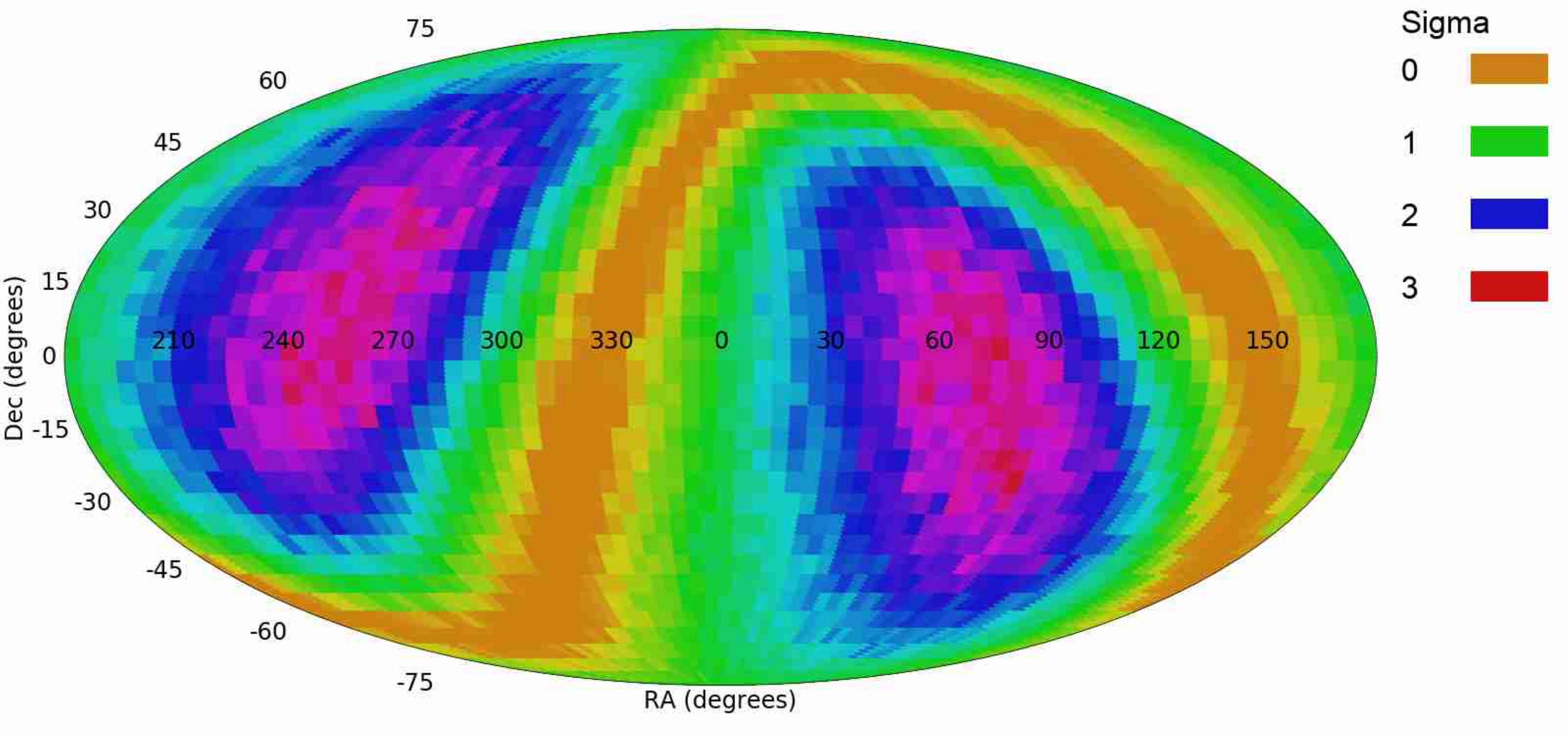}
\caption{Statistical signal of a dipole axis formed by galaxy spin directions in different parts of the sky \citep{mcadam2023reanalysis}.}
\label{longo_z_results}
\end{figure}

The third experiment that \cite{petal} tried to reproduce is \citep{longo2011detection}. \cite{petal} argued that the experiment was ``a shot in the dark'' since no clear explanation of the statistical method was found in \citep{longo2011detection}. This claim is surprising, as the explanation in \cite{longo2011detection}
is sufficiently clear, and this author had no difficulty understanding and implementing it to reproduce the results. The analysis showed a dipole axis with statistical significance of 2.92$\sigma$. This is somewhat weaker than the 3.16$\sigma$ reported in \cite{longo2011detection}, but the difference is not dramatic, and the analysis should be considered reproducible. Figure~\ref{longo_z_results} shows the statistical significance that an axis exists in each part of the sky. The results are in close alignment with a large-scale cosmological axis that peaks at $(\alpha=217^o,\delta=32^o)$ \citep{longo2011detection}, and the experiment can be considered reproducible. 

It is difficult to understand what led \cite{petal} to claim that these three experiments are not reproducible, and why the replication of the analysis led them to different results than stated in the papers. Also, code was available at the same web page from which one of the datasets was downloaded \url{https://people.cs.ksu.edu/~lshamir/data/iye_et_al}. Given that the code is provided with clear step-by-step instruction of how to use it, it is unclear why \cite{petal}chose not to use that code. But even without a pre-written code, the reproduction is simple, and the results show excellent agreement with the results reported in the papers. An immediate explanation is a code error. Interestingly, \cite{petal} do not specify the results of their attempts to reproduce the experiments, but merely state that they did not match the information stated in the paper. They do not indicate whether their results were statistically significant, or provide any other information, making it more difficult to understand the specific nature of the coding error. The code used by \cite{petal} is not publicly available or accessible in any other form, and therefore there is no practical way to identify the specific error that prevented  \cite{petal} from reproducing the experiments correctly.


\section{Conclusion}
\label{conclusion}
 
Confirmation bias is a common and well-documented phenomenon according which people take substantial actions to confirm their existing beliefs \citep{hart2009feeling}. Confirmation bias has been shown to be extremely dominant in science, and exhibits itself in science in the form of resisting new discoveries by selectively interpreting data or designing experiments in a manner that would lead to their favorable results \citep{nickerson1998confirmation}, or ignoring and misinterpreting studies that do not conform to current beliefs \citep{letrud2019affirmative,koehler1993influence,mahoney1977publication,oswald2004confirmation}. 

The direction towards which a galaxy rotates is a feature of the perspective of the observer, and therefore the null hypothesis is that the galaxy spin directions as observed from Earth are distributed randomly. Previous studies that examined the distribution of galaxy spin directions have shown elements that can be associated with confirmation bias \citep{sym15091704}. For instance, the experiment using ``Galaxy Zoo'' crowdsourcing annotations argued that a large number of galaxies show that galaxy spin directions are distributed randomly \citep{land2008galaxy}. But the vast majority of the galaxies used in that study were annotated manually without mirroring the images to offset for the human bias, leading to a very strong bias. The conclusion of the authors was therefore not based on the large number of galaxies in Galaxy Zoo, but on a far smaller dataset in which the galaxy images were mirrored, which was created after the problem of human bias was noticed. Also, the paper states that the smaller dataset agreed with the null hypothesis, yet without providing a statistical inferebce or a $p$ value. Statistical analysis of that same data showed very good agreement in both direction and magnitude with non-random distribution \citep{sym15091704}.

\cite{hayes2017nature} used computer analysis to annotate the galaxy images, which lead to statistically significant non-random distribution in most of their experiments. The one experiment that showed no statistically significant signal was done by applying a step of machine learning classification of spiral galaxies such that all features that correlate with the galaxy directions of rotation were removed manually. The paper does not state the reason for manually removing these features. As shown in \citep{mcadam2023reanalysis}, removal of these features led to random distribution. The removal of these features led to the desired outcomes that are also stated in the abstract, but careful reading of the paper shows an unexpected practice that is not explained in their paper \citep{mcadam2023reanalysis}.

\cite{iye2020spin} applied a 3D analysis method to show that the galaxy spin directions are distributed randomly. Their analysis required that each galaxy has its redshift, and the paper reports using the ``measured redshift'' of each galaxy. That claim is challenged by the fact that the vast majority of these galaxies did not have spectra. A careful reading of the journal version of the paper reveals a citation of a photometric redshift catalog. Clearly, the high inaccuracy of the photometric redshift is expected to weaken the statistical signal.  The citation of the photometric redshift catalog does not exist in the arxiv version, and a person reading the arxiv version cannot know about the use of the photometric redshift even after careful reading of the paper and all of its references \citep{psac058}. \cite{iye2020spin} also made the assumption that the galaxies are uniformly distributed in the hemisphere, which is not required since all galaxies have their exact locations, and is also crudely incorrect in the case of SDSS galaxies. That assumption is not mentioned in the paper, and one needs to contact the institution where the research was conducted (NAOJ) to learn that uniform distribution of the galaxies in the hemisphere was assumed \citep{sym15091704}.

Interestingly, previous to that the same authors also analyzed a large dataset of galaxies in a relatively small footprint, and found strong statistically significant non-random distribution \citep{tadaki2020spin}. Yet, they chose to highlight the null hyphothesis, stating that there is no statistically significant non-random distribution \citep{tadaki2020spin}. The analysis method indeed had a certain error, but the results are in far better agreement with non-random distribution than with the null hypothesis \citep{sym15091704}.

\cite{petal} argue that the distribution of galaxy spin directions in the Universe observed from Earth is random. Their argument is based on the contention that the probability of a galaxy to rotate in a certain direction does not follow normal distribution. While that argument can be considered challenging, it is complemented by the claim that previous experiments using the straightforward and previously tested $chi^2$ method \citep{sym15091704} were not reproducible, and did not lead to the same results shown in the papers. Therefore in either case the null hypothesis is correct. That is, whether assuming normal distribution or not assuming normal distribution, in both cases the null hypothesis of random distribution is supported by the results of the experiments.

As stated in Section~\ref{reproducability}, the previous experiments \citep{mcadam2023reanalysis,longo2011detection} are fully reproducible. As shown in Section~\ref{empirical}, empirical results show that the new method proposed by \cite{petal} is not consistent with the data, and does not respond even in cases where the distribution of the data is extremely non-random. Code for the new method is not available, making it difficult to know what could lead to such inconsistency, but the unsuccessful attempt to reproduce experiments that can easily be reproduced as discussed in Section~\ref{reproducability} provides an indication of a coding error.

The work of \cite{petal} also shows evidence of confirmation bias. For instance, when analyzing the possible impact of the redshift,  \cite{petal} chose to use the \citep{longo2011detection} dataset, which is far smaller than the larger datasets, and therefore less likely to provide statistical significance. \cite{petal} explain that the reason for the selection is that the catalog of \citep{longo2011detection} is the only catalog that its galaxies have spectroscopic redshift. That argument is obviously not correct, as the far larger datasets used in \citep{mcadam2023reanalysis} and also used by \cite{petal} are based on the original Galaxy Zoo galaxies, all of them have spectra. These catalogs are an order-of-magnitude larger than \citep{longo2011detection}.


With the exception of \citep{longo2011detection}, the datasets studied by \cite{petal} were not used to claim for the existence of a cosmological-scale dipole axis, but to compare different results and identify the reasons for the inconsistency in the conclusions of different studies. Previous papers that used these datasets \citep{psac058,mcadam2023reanalysis} stated clearly that these datasets cannot be considered a proof for the existence of a dipole axis. Therefore,  even if the experiments made by \cite{petal} were reproducible and correct, they do not support the conclusion.

The unsuccessful attempts to reproduce several previous experiments can also be the result of confirmation bias, as the reproduction is technically simple, and the results reported in the papers can be reproduced with very high accuracy. The title of \citep{petal} might also be biased by the null hypothesis, as none of the datasets analyzed by \cite{petal} were used to claim for the presence of a dipole axis, with the exception of \citep{longo2011detection}. It is surprising, however, that the small dataset of the Panoramic Survey Telescope and Rapid Response System (Pan-STARRS) galaxies showed a statistically significant dipole axis. These results can be considered in conflict with the abstract, stating that ``All analysis indicate consistency with isotropy to within 3$\sigma$''. \cite{petal} also do not refer to observations with a high number of galaxies that do not require a statistical fitting of the data into a pre-defined model, as shown in Figure ~\ref{all10_6K_dist90_inc1} \citep{shamir2022analysis}.

As discussed in Section~\ref{introduction}, numerous studies using multiple statistical analysis methods show non-random distribution in galaxy spin directions as observed from Earth \citep{longo2011detection,shamir2012handedness,shamir2016asymmetry,shamir2019large,shamir2020patterns,shamir2020large,shamir2020pasa,shamir2021particles,shamir2021large,shamir2022new,shamir2022large,shamir2022asymmetry,shamir2022analysis,Shamir_2024}. These observations are aligned with multiple other probes and analyses that show that the Universe as observed from Earth is not isotropic \citep{Aluri_2023}.

Another possibility is that the asymmetry is not a feature of the large-scale structure of the Universe, but a feature of the physics of galaxy rotation. Namely, the rotational velocity of the observed galaxies relative to the rotational velocity of the Milky Way affects the brightness of the galaxies in a manner that is beyond the expected Doppler shift effect given the $\sim220km\cdot sec^{-1}$ rotational velocity of the Milky Way \citep{shamir2020asymmetry,sym15061190}. That is, due to the Doppler shift effect galaxies that rotate in the same direction relative to the Milky Way are expected to have a slightly different brightness than galaxies that rotate in the opposite direction relative to the Milky Way. If galaxies rotating in certain directions around the galactic pole are brighter, it means that more of these galaxies will be seen from Earth, leading to a difference in the number of galaxies that rotate in opposite directions. Such difference is expected to peak at the galactic poles, which is indeed observed empirically as briefly discussed in the end of Section~\ref{normal_distribution}. 
Given the rotational velocity of the Milky Way, the brightness differences is expected to be low \citep{sym15061190}. But it should be reminded that the physics of galaxy rotation is still one of science's greatest mysteries. The common explanation of dark matter has not yet been proven, and there is a body of evidence that challenge the correctness of that theory \citep{kroupa2012dark}. It is therefore should not be ruled out that the physics of galaxy rotation exhibits itself in an unexpected form, and corresponds physically to a far higher velocity. The spectroscopy and photometry of galaxies observed from Earth correspond to velocity that is roughly five to 10 times faster than the rotational velocity of the Milky Way \citep{shamir2024simple,shamir2020asymmetry,sym15061190}.  

The contention that the probability of a galaxy to rotate in a certain direction does not follow normal distribution \citep{petal} is a challenging claim. This paper shows that it leads to numerous inconsistencies, and is in disagreement with the properties of the data it attempts to analyze, including simple Monte Carlo simulations. Evidence of coding errors can also be identified in the attempts to reproduce previous experiments. The absence of open source code prevents from fully understanding the nature of the inconsistencies and the reasons leading to them. That further reinforces the need to have an open source code practice in astronomy \citep{shamir2013practices}. It also emphasizes the need to test statistical methods in open forums, and in full transparency. But given the large number of observations, including numerous instruments and different analysis methods, it is required to investigate the reasons for the abnormal distribution of galaxy spin directions as observed from Earth.

\bibliographystyle{apalike}
\bibliography{main}

\end{document}